\documentclass[fleqn,twoside]{article}
\usepackage{espcrc2}
\usepackage{graphicx,amssymb,lineno}

\begin{document}
\title{The construction of the CMS Silicon Strip Tracker}

\author{Giacomo Sguazzoni\address[INFN]{INFN Sezione di Firenze, via Sansone, 1 - I-50019 Sesto F.no (FI) - ITALY} on behalf of the CMS Silicon Strip Tracker Collaboration%
}

\begin{abstract}
The CMS Silicon strip tracker is a very large scale tracker entirely based on Silicon strip detector technology. It is build up of $\sim$15k Silicon strip modules for a total of $\sim$9M analogue readout channels with an overall active silicon area of $\sim$200m$^2$, to be operated at -10$^\circ$C and able to survive for 10 years to the LHC radiation environment. The integration of modules, electronics, mechanics and services has been completed within the last two years; large standalone sub-structures (shells, disks, rods and petals depending on the tracker subdetector) have been first integrated and verified; then they have been brought together into the final configuration.
     The CMS silicon tracker design and its construction is reviewed with particular emphasis on the procedures and quality checks deployed to successfully assembly several modules and all ancillary components into these large sub-structures. An overview of the results and the lesson learned from the tracker integration are given, also in terms of failure and damage rates.
\end{abstract}



\maketitle

\section{Introduction}

The CMS tracker is the world largest Silicon strip detector with its $\sim$$23$m$^3$ instrumented by 15148 Silicon strip modules for a total 198m$^2$ of Silicon active area and  9.316.352 channels with full optical analog readout, to be operated at $-10^\circ$C and resistant within the LHC radiation environment~\cite{Sguazzoni:2004sx}. 

\begin{figure}[t]
\begin{center}
\includegraphics*[width=0.45\textwidth]{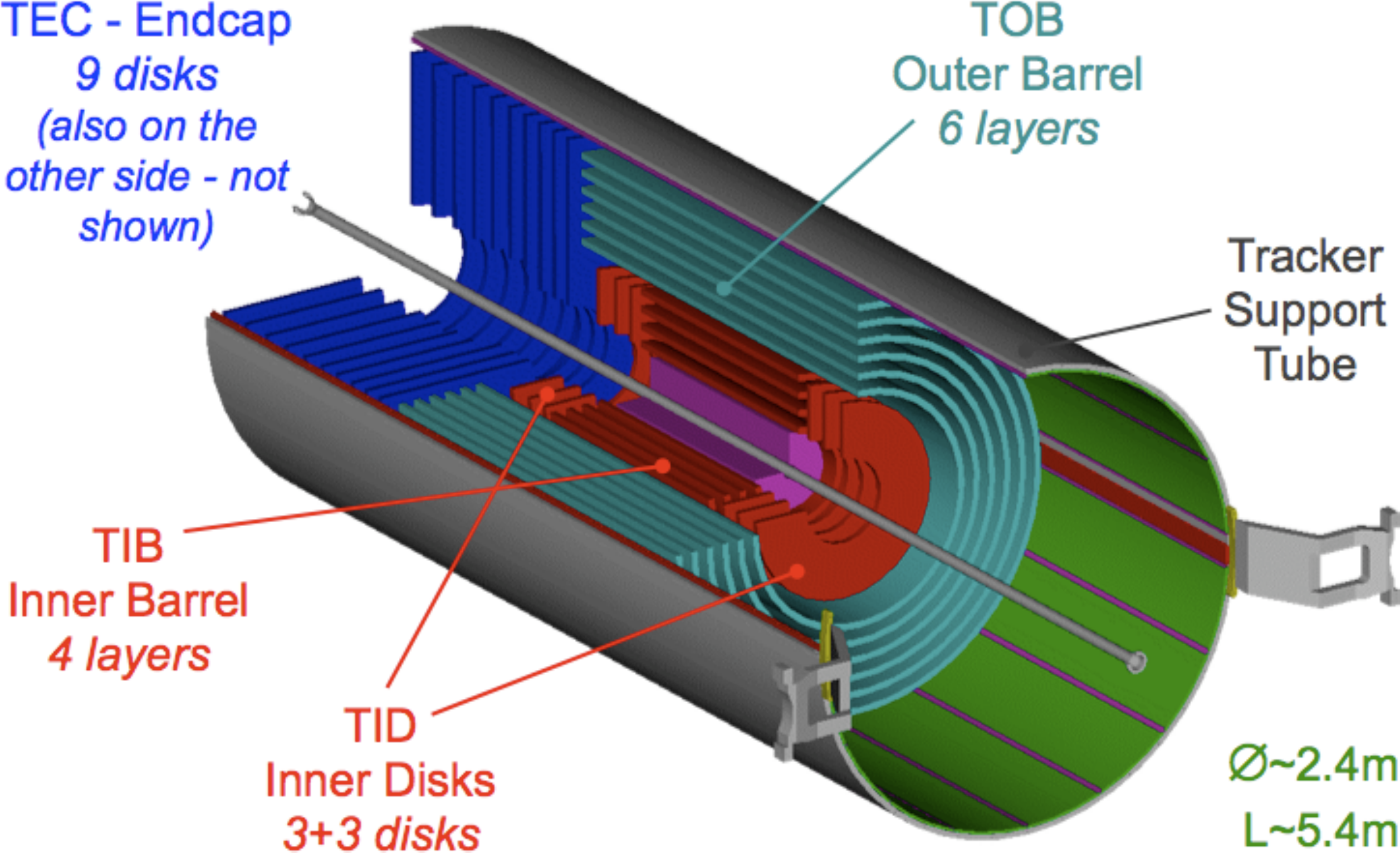}
\end{center}
\vskip -5mm
\caption{A sketch of the CMS Silicon strip tracker.}
\vskip -5mm
\label{fig:sst}
\end{figure}

The Silicon Strip Tracker (SST), sketched in Fig.~\ref{fig:sst}, occupies the radial range  between 20cm and 110cm around the LHC interaction point. The barrel region ($| z |  < 110$cm) is split into a Tracker 
Inner Barrel (TIB) made of four detector layers, and a Tracker Outer Barrel (TOB) made of six detector layers. The TIB is shorter then the TOB, and is complemented by three Tracker Inner Disks per side (TID) each made of three rings. The forward and backward regions $120{\rm cm} < |z| < 280 {\rm cm}$ are covered by nine Tracker End-Cap (TEC) disks per side, each made of four to seven rings.

A high modularity has been the design cornerstone to ease the detector construction from the massive production of Si-strip modules~\cite{Sguazzoni:2004zt} and all other components. This has been made possible by building small but fully self-sufficient carbon fiber assemblies supporting several modules together with cooling and readout electronics. Easier handling and mounting merges with effective quality assurance and control. All substructures have been tested for basic functionality (pedestal and noise), cooling robustness and, later, for more complex functionality (system tests and track reconstruction).

\section{Quality control tests on substructures}

\begin{figure}[t]
\begin{center}
\includegraphics*[width=0.41\textwidth]{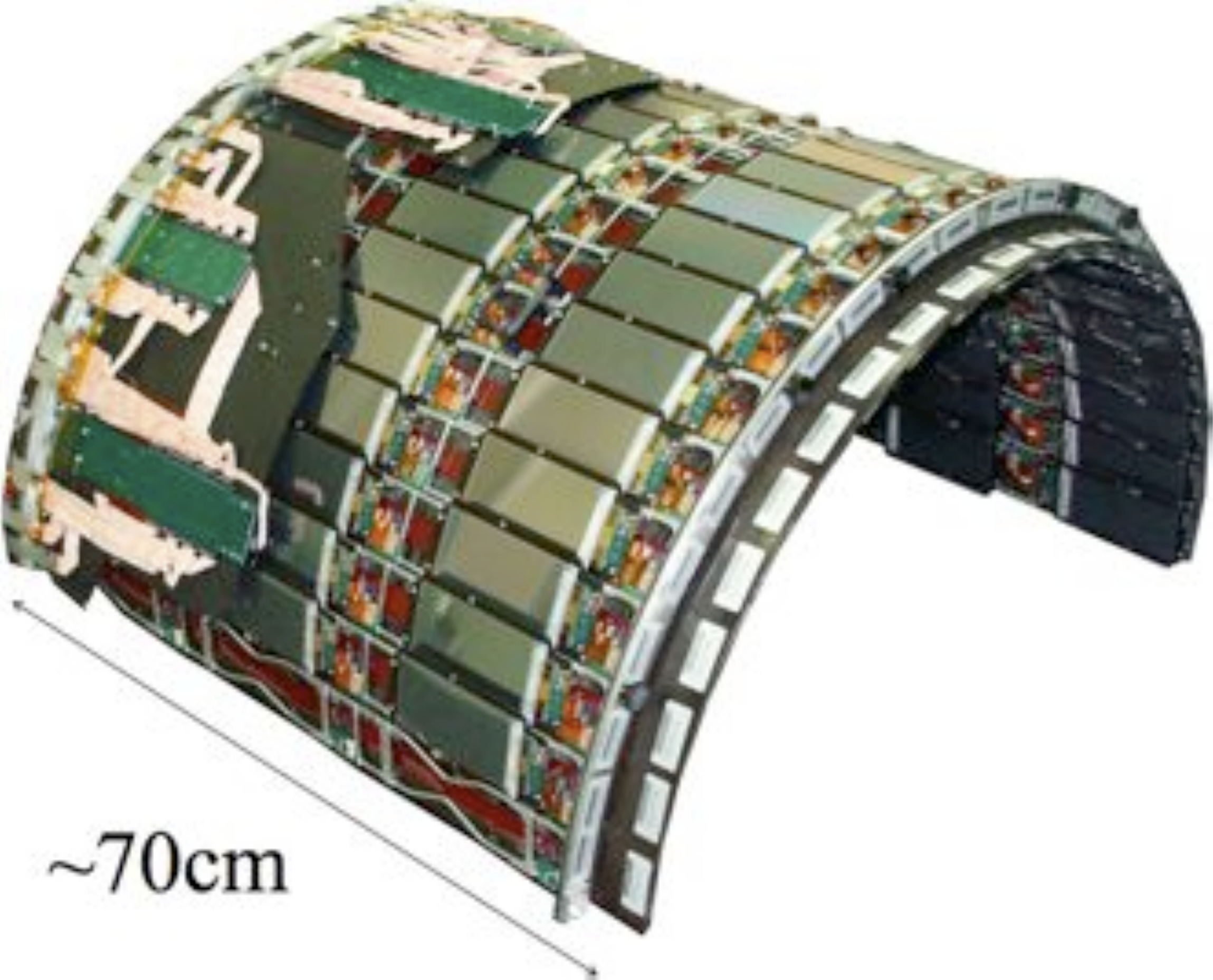}
\end{center}
\vskip -9mm
\caption{A shell of the third layer of the TIB.}
\vskip -3mm
\label{fig:shell}
\end{figure}

\begin{figure}[t]
\begin{center}
\includegraphics*[width=0.29\textwidth]{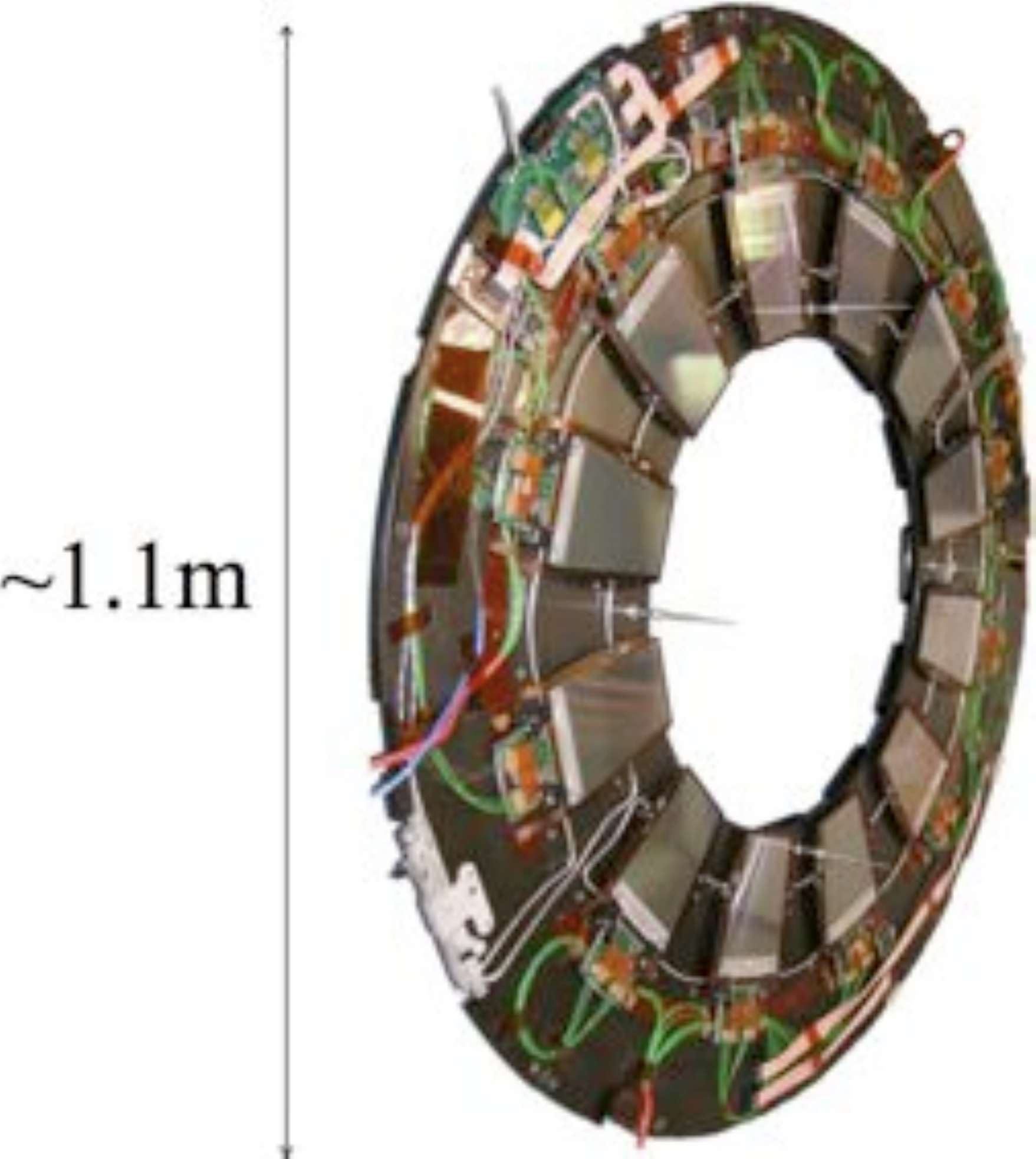}
\end{center}
\vskip -9mm
\caption{The innermost ring of a TID disk.}
\vskip -3mm
\label{fig:ring}
\end{figure}

\begin{figure}[t]
\begin{center}
\includegraphics*[width=0.44\textwidth]{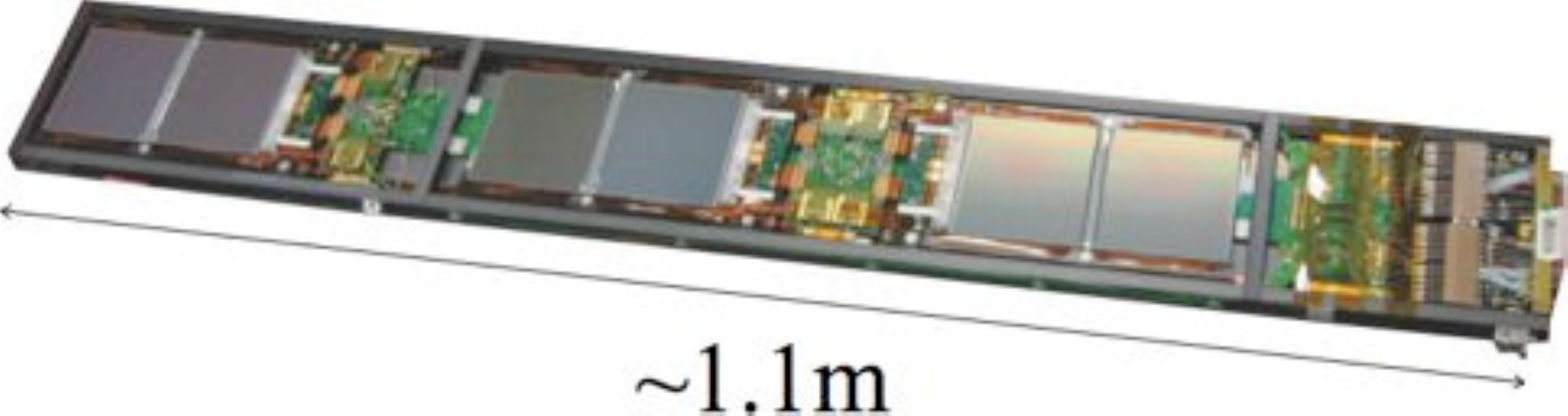}
\end{center}
\vskip -9mm
\caption{A rod of the TOB.}
\vskip -3mm
\label{fig:rod}
\end{figure}

\begin{figure}[t]
\begin{center}
\includegraphics*[width=0.39\textwidth]{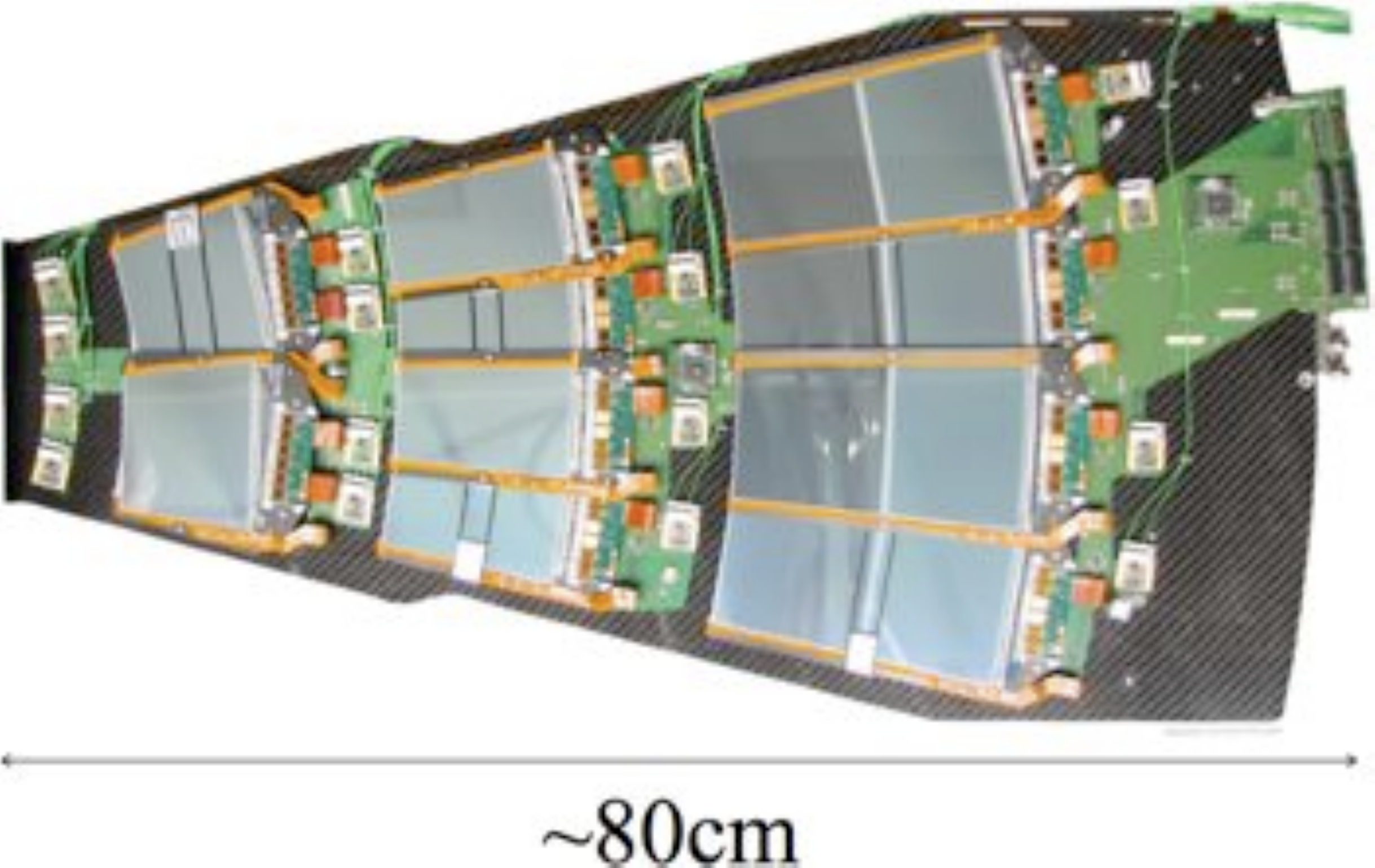}
\end{center}
\vskip -9mm
\caption{A petal of the TEC.}
\vskip -3mm
\label{fig:petal}
\end{figure}

The following subdetector dependent substructures can be found in the tracker: the TIB is split into 16 half cylinder {\em shells} (Fig.~\ref{fig:shell}) that host 135 to 216 modules; the TID is structured in $3$ {\em rings} (Fig.~\ref{fig:ring}) per disk, each designed to support 40 or 48 modules; the TOB is made of 688 {\em rods} (Fig.~\ref{fig:rod}), sort of drawers providing support for 6 or 12 modules; the TEC are build up of $144$ {\em petals} (Fig.~\ref{fig:petal}) per endcap, each hosting 17 to 28 modules.

During assembly steps, each single component is verified. Optical links are checked by measuring the output level of an auxiliary signal issued by the readout chips and modules are checked by pedestal and noise runs at full bias (400V) to spot high voltage issues and bad channels~\cite{cc,smersi}.

Once substructures are completed, system wide tests are performed by providing cooling as in final CMS operating conditions. Temperature probes hosted on modules are effective to spot cooling circuit problems~\cite{palla}.
A cosmic ray setup allows to perform track reconstruction exercises, to measure S/N ratio for MIPs and to check the overall mechanical precision verified to be at the sub-millimeter level.

\section{Conclusions}

The SST was completed in March 2007 and will be installed in CMS within the summer. To now, the resulting total fraction of bad channels is 0.21\%: 0.07\% due to module failures, 0.05\% due to bad optical links and 0.09\% of sparse defects, mostly preexistent on modules before the integration. The assembly procedures and the stringent quality control tests have been proved to be sound and effective.

\end{document}